\documentclass[11pt,a4paper]{article}
\usepackage{epsfig}
\usepackage{amsfonts}
\usepackage{rotating}
\usepackage{cite}

\def\3{\ss}

\newcommand{\Coll}{Collaboration}

\newcommand{\la}{\langle}
\newcommand{\ra}{\rangle}
\newcommand{\bec}{\begin{center}}
\newcommand{\eec}{\end{center}}
\newcommand{\beq}{\begin{equation}}
\newcommand{\eeq}{\end{equation}}

\def\as{\alpha_s}
\def\va{\varepsilon}
\def\T{\Theta}
\def\t{\theta_{12}}
\def\nc{n_{\mathrm{c}}}
\def\nt{n_{\mathrm{t}}}

\def\sc{\sigma_{\mathrm{c}}}
\def\st{\sigma_{\mathrm{t}}}

\def\Nev{N_{\mathrm{ev}}}

\def\cov{\mathrm{cov}(\nc , \nt)} 
\def\GeV2{\mathrm{GeV}^2}
\def\E{e^+e^-}
\def\pp{pp}
\def\ppb{p\overline{p}}
\def\p+p{\pi^{\pm}p}
\def\K+p{K^{+}p}
\def\m+p{\mu^{+}p}
\def\np{\nu p}
\def\bnp{\overline{\nu}p}
\def\ep{e^+p}

\def\YH{\hat{Y}(\varepsilon )}
\def\Y{Y(\varepsilon )}  

\begin{document}
 
\title 
{
\begin{flushright}{\large DESY--99--063 journal version}\end{flushright} 
\vspace{2cm}
\bf\Large   Angular and Current-Target Correlations in  \\ 
            Deep Inelastic Scattering at HERA\\
\vspace{1cm}}

\author{ZEUS Collaboration }
\date{}
\maketitle
\begin{abstract}
\noindent
Correlations between  charged particles in deep inelastic 
$\ep$ scattering have been studied in the
Breit frame with the ZEUS detector at HERA using
an integrated luminosity of 6.4 pb$^{-1}$.   
Short-range correlations are analysed in terms
of the angular separation 
between current-region  particles 
within a cone centred around the virtual photon axis. 
Long-range correlations   
between the current and target 
regions have also been measured.
The data support predictions for the scaling behaviour of the angular
correlations at high $Q^2$ and for anti-correlations 
between the current and target regions over a large range in $Q^2$ and
in the Bjorken scaling variable $x$.
Analytic QCD calculations and Monte Carlo models correctly
describe the trends of the data at high $Q^2$, but show
quantitative discrepancies.
The data  show  differences  between 
the correlations in deep inelastic scattering 
and  $\E$  annihilation.
\end{abstract}

\pagestyle{plain}
\thispagestyle{empty}
\clearpage
\pagenumbering{Roman}
 
\begin{center}                                                                                     
{                 \Large  The ZEUS Collaboration              }                               
\end{center}                                                                                       
  J.~Breitweg,                                                                                     
  S.~Chekanov,                                                                                     
  M.~Derrick,                                                                                      
  D.~Krakauer,                                                                                     
  S.~Magill,                                                                                       
  B.~Musgrave,                                                                                     
  A.~Pellegrino,                                                                                   
  J.~Repond,                                                                                       
  R.~Stanek,                                                                                       
  R.~Yoshida\\                                                                                     
 {\it Argonne National Laboratory, Argonne, IL, USA}~$^{p}$
% ...
\newpage                                                                                           
$^{\    1}$ now visiting scientist at DESY \\                                                      
% ...
$^{  36}$ partially supported by the Foundation for German-Russian Collaboration                   
DFG-RFBR \\ \hspace*{3.5mm} (grant no. 436 RUS 113/248/3 and no. 436 RUS 113/248/2)\\              
% ...
                                                           %
                                                           %
\newpage   
                                                           %
                                                           %
\begin{tabular}[h]{rp{14cm}}                                                                       
$^{a}$ &  supported by the Natural Sciences and Engineering Research                               
          Council of Canada (NSERC)  \\                                                            
$^{b}$ &  supported by the FCAR of Qu\'ebec, Canada  \\                                            
$^{c}$ &  supported by the German Federal Ministry for Education and                               
          Science, Research and Technology (BMBF), under contract                                  
          numbers 057BN19P, 057FR19P, 057HH19P, 057HH29P, 057SI75I \\                              
$^{d}$ &  supported by the MINERVA Gesellschaft f\"ur Forschung GmbH, the                          
German Israeli Foundation, and by the Israel Ministry of Science \\                                
$^{e}$ &  supported by the German-Israeli Foundation, the Israel Science                           
          Foundation, the U.S.-Israel Binational Science Foundation, and by                        
          the Israel Ministry of Science \\                                                        
$^{f}$ &  supported by the Italian National Institute for Nuclear Physics                          
          (INFN) \\                                                                                
$^{g}$ &  supported by the Japanese Ministry of Education, Science and                             
          Culture (the Monbusho) and its grants for Scientific Research \\                         
$^{h}$ &  supported by the Korean Ministry of Education and Korea Science                          
          and Engineering Foundation  \\                                                           
$^{i}$ &  supported by the Netherlands Foundation for Research on                                  
          Matter (FOM) \\                                                                          
$^{j}$ &  supported by the Polish State Committee for Scientific Research,                         
          grant No. 115/E-343/SPUB/P03/154/98, 2P03B03216, 2P03B04616,                             
          2P03B10412, 2P03B05315, and by the German Federal Ministry of                            
          Education and Science, Research and Technology (BMBF) \\                                 
$^{k}$ &  supported by the Polish State Committee for Scientific                                   
          Research (grant No. 2P03B08614 and 2P03B06116) \\                                        
$^{l}$ &  partially supported by the German Federal Ministry for                                   
          Education and Science, Research and Technology (BMBF)  \\                                
$^{m}$ &  supported by the Fund for Fundamental Research of Russian Ministry                       
          for Science and Edu\-cation and by the German Federal Ministry for                       
          Education and Science, Research and Technology (BMBF) \\                                 
$^{n}$ &  supported by the Spanish Ministry of Education                                           
          and Science through funds provided by CICYT \\                                           
$^{o}$ &  supported by the Particle Physics and                                                    
          Astronomy Research Council \\                                                            
$^{p}$ &  supported by the US Department of Energy \\                                              
$^{q}$ &  supported by the US National Science Foundation \\                                       
\end{tabular}                                                                                      
                                                           %
                                                           %

\newpage
\setcounter{page}{1}
\pagenumbering{arabic}

\section{Introduction}
\label{sec:intr}

This paper reports  a  study of both short-range 
and long-range correlations in neutral-current
deep inelastic  $\ep$ scattering events 
measured  at a centre-of-mass energy
of $\sqrt{s}=300$ GeV with the ZEUS detector at HERA.
   
In deep inelastic  scattering (DIS) at HERA energies, final-state
charged particles have been studied  
in terms  of  fragmentation functions \cite{zeusM,zeusFF,h1FF,zeusMN},  
transverse momentum spectra \cite{zeusMT,trH1} and      
particle correlations \cite{H1c1,H1c2,H1c3}. 
The short-range correlations between final-state particles at  small
phase-space separations  are sensitive to  the structure of
many-particle inclusive densities \cite{rew}. 
The study of long-range correlations 
can aid the understanding of the interdependence
between distant phase-space regions;  these measurements   complement
and extend 
the studies of the global multiplicity distributions  that  have already
been reported  in DIS \cite{zeusM,h1FF,h1M} at HERA energies. 

Two-particle  angular 
correlations have recently been studied in $\E$ collisions 
by the DELPHI Collaboration \cite{cor_delphi} at LEP and  
many-particle  correlations have also been measured  by 
the L3 \cite{fluc_l3} and OPAL 
Collaborations \cite{fluc_opal}. Such  studies   
provide  more detailed insight  into the 
QCD models (see reviews \cite{QCD,QCDh}) based on 
the hypothesis of Local Parton-Hadron Duality, 
which states   
that the single-particle distributions of hadrons are
proportional  to  the corresponding parton spectra \cite{AZIM}. 
The analytic predictions, derived in the 
Double Log Approximation (DLA)
\cite{theo_cor,theo_fluc1,theo_fluc2},  have shown 
discrepancies with the observed angular correlations  as well as with  
the fluctuation  measurements \cite{cor_delphi,fluc_l3}. 
The DLA calculations  take into account the contributions from the
infrared and collinear singularities of the gluon emissions
and  the  angular ordering (colour coherence), but neglect   
energy-momentum conservation in gluon splittings and 
$q\bar{q}$ production as well as  finite-energy 
effects in the QCD parton cascade. 
However, Monte Carlo (MC) models which take 
these effects into account  and which, in addition,
contain  hadronisation followed 
by resonance decays, also
show discrepancies with the data \cite{fluc_l3,fluc_opal}.  

DIS provides a unique possibility to confront both the  analytic results
and MC models with data at various energy scales. 
The main motivation for such studies is to investigate
the applicability of the perturbative analytic calculations 
to DIS data at HERA energies. 
The two-particle angular correlations in the 
current region of the Breit frame 
are also compared with the  correlations in a single hemisphere of 
$\E$ annihilation in order to study a possible universality 
of two-particle spectra in these processes.   
Recent comparisons of single-particle spectra  
have shown  considerable similarity 
between DIS at high $Q^2$ and $\E$ annihilation 
\cite{zeusM, zeusMN, zeusFF, h1FF, h1M}.  

Correlations between
distant phase-space regions provide important information
on multi-hadron production that cannot be studied by 
measuring observables separately in each phase-space region.  
In DIS, a measurement of  the correlations between 
the hemisphere populated  by hadrons originating from the struck quark  
and that containing  predominantly particles from the proton remnant   
can be made in a way analogous to ``forward-backward'' correlations.
Such long-range correlations in rapidity 
have been studied for many years \cite{ee,mp,del,al,op,np,pp,Kp,H1c3}. 
Results for $\E$ and $\m+p$ processes
indicated  that such correlations are small at  
low energies \cite{ee,mp}. At LEP1 energies, 
positive long-range correlations were observed, mainly
due to heavy quark pair production \cite{del,al,op}.
For $\np$ and $\bnp$ processes, the correlations are
small and negative \cite{np}.
For $\pp$, $\ppb$ \cite{pp} and  $\p+p$,
$\K+p$ \cite{Kp} collisions,
the correlations are  positive and
increase  with $\sqrt{s}$. Recently,  the H1 Collaboration has   shown
that the forward-backward correlations measured in the
$\gamma^*$-pomeron 
centre-of-mass system of
diffractive $\ep$ collisions are positive \cite{H1c3}.
In contrast, in the Breit frame in DIS,   
negative  long-range correlations  due to the kinematics of 
first-order QCD effects have been predicted \cite{ch}.  

\section{Definitions and analytic QCD results}
\label{sec:meth}
\subsection{DIS  kinematics in the Breit frame}

The event kinematics  of  DIS  processes are  
determined by the negative squared 4-momentum transfer
at the lepton vertex,   
$Q^2=-q^2=-(k-k^{'})^2$ ($k$ and $k^{'}$ denote the 4-momenta of the
initial and final-state positron, respectively), 
and the Bjorken
scaling variable $x=Q^2/(2 P_{\mathrm{p}}\cdot  q)$, 
where $P_{\mathrm{p}}$ is the
4-momentum of the proton.
The fractional energy transfer   
to the proton in its rest frame, $y$,    
is related to these two variables
by  $y\simeq Q^2/xs$,
where $\sqrt{s}$ is the positron-proton
centre-of-mass energy.
                                                         
\begin{figure}
\begin{center}
\vspace{1.0cm}
\mbox{\epsfig{file=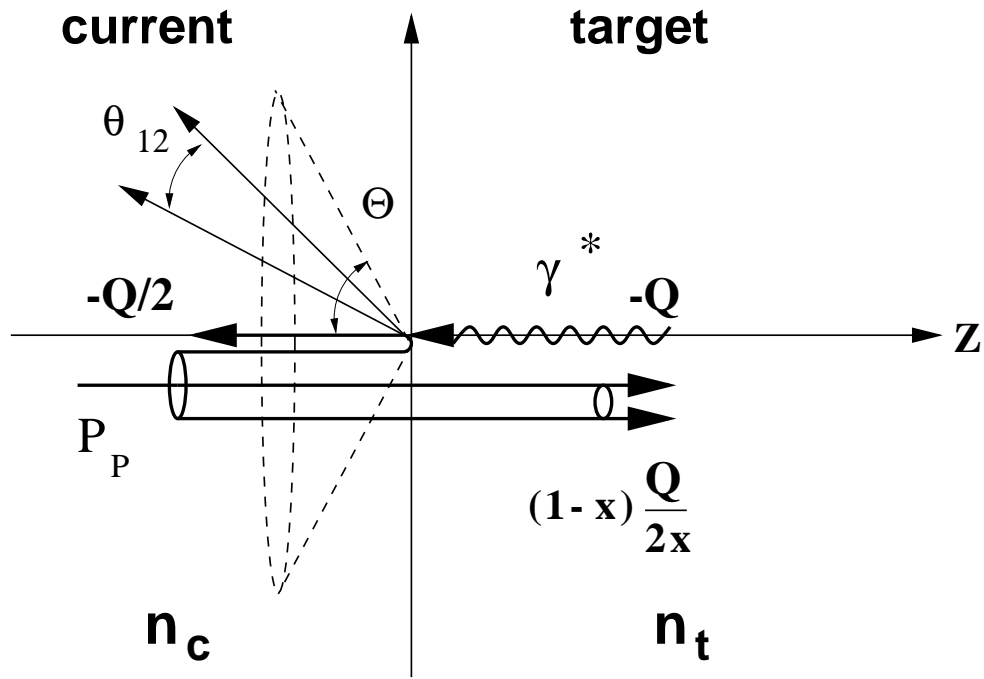,height=9.0cm}}
\caption{ 
A schematic representation of  the measurement of
correlations in the Breit frame. The angular  correlations in the
current region 
are measured  between any two charged particles separated by $\t$. 
Long-range correlations in  the full phase space
are measured between current ($\nc$) and target
($\nt$) multiplicities.}  
\label{fig:1}
\end{center}
\end{figure}

The correlations are studied in the Breit
frame \cite{Br}  which provides a natural system to separate
the radiation from  the outgoing
struck quark from  that associated with the rest of the
hadronic final state. In this frame, 
the exchanged virtual boson is completely
space-like and has a momentum $q=(q_0,q_X,q_Y,q_Z)=(0,0,0,-Q)$ 
as shown in Fig.~\ref{fig:1}. For the Quark-Parton Model (QPM), the 
incident quark carries $Z$-momentum $p_Z^{\mathrm{Breit}}=Q/2$ 
in the positive $Z$-direction and the outgoing
struck quark carries $p_Z^{\mathrm{Breit}}=-Q/2$ 
in the negative $Z$-direction. 
The phase space of the event can be divided into two
regions. All particles with negative $p_Z^{\mathrm{Breit}}$ 
form the current region. These  particles   
are produced by the fragmentation   of
the struck quark, so that the current region is analogous to a
single hemisphere of  $\E$ annihilation.  
Particles with positive $p_Z^{\mathrm{Breit}}$  are
assigned to the target region associated with the incoming proton.
The longitudinal phase space in $p_Z^{\mathrm{Breit}}$ 
available for the target-region particles extends to  
$p_Z^{\mathrm{Breit}}=(1-x)Q/2x$. This means that 
the current and target regions
are highly  asymmetric at small $x$. Thus the Breit frame 
differs  from the hadronic centre-of-mass $(\gamma^* p)$ frame and
the centre-of-mass frame of $\E$ annihilation.  
 
This simple picture of DIS in the Breit frame is changed when
leading-order QCD processes, Boson-Gluon Fusion (BGF) 
and QCD-Compton (QCDC) scattering, are considered.
In such processes, the collision 
in the Breit frame is no longer collinear, although
the current and target  regions are still well-defined
operationally \cite{Str}. 

\subsection{Two-particle  correlations}

The correlations between charged particles in the current region 
are studied using the angle between 
any two particles (see Fig.~\ref{fig:1}).
For each pair of particles 
inside a cone of half-angle $\Theta$  centred around  the Breit frame axis, 
the relative angle $\theta_{12}$ is determined.
For a given  QCD scale, $\Lambda$, 
this angle is transformed into the variable 
\beq
\va =\frac{\ln(\T/\t )}{\ln(P\sin\T /\Lambda)},  
\label{1}
\eeq
For DIS in the Breit frame, $P=Q/2$ is chosen,  corresponding to      
the outgoing quark momentum in the QPM. Note
that $\Lambda$ is an effective scale parameter which is 
not related to  $\Lambda_{\rm \overline{MS}}$ \cite{cor_delphi}.  
The two-particle inclusive density  
\beq 
\rho (\va )=\frac{1}{\Nev} \frac{dn_{\mathrm{pair}}}{d\va} 
\label{1a}
\eeq 
is determined, where $\Nev$ is  the number of events and 
$(dn_{\mathrm{pair}}/d\va)\delta\va $ is the number of particle pairs
in the interval $\va $ to $\va + \delta\va$.
The definition (\ref{1})  is used  
since the DLA predictions can be naturally  
expressed in this scaling  variable \cite{theo_cor, theo_fluc2} and 
the logarithmic dependence of (\ref{1}) expands  small values of $\t$.    
Note that our definition of $\va$  is  different from the 
scaling variable with the denominator $\ln (P\T/\Lambda)$ 
used previously \cite{cor_delphi, fluc_l3}:  
for large $\T$, the definition (\ref{1}) is better suited to
compare the DLA predictions with the data \cite{WOchs}. 
  
In this paper  we  measure    
the  relative angular distribution $\hat{r}(\va )$
and the  correlation function $r(\va )$:
\beq 
\hat{r}(\va )=\frac{\rho(\va )}{\la n  (\T ) \ra^2}, \qquad
r(\va ) = \frac{\rho(\va )}{\rho_{\mathrm{mix}}(\va )}. 
\label{3} 
\eeq
The  variable $\hat{r}(\va )$ is  $\rho(\va )$
normalised using  the average   
charged multiplicity $\la n  (\T ) \ra$ in the cone.
Even for uncorrelated particle production, 
$\hat{r}(\va )$ depends on $\va$ in a manner 
which is determined by the single-particle spectra. 
This dependence is reduced in $r(\va )$,  
which is normalised instead by $\rho_{\mathrm{mix}}(\va )$, 
calculated using particles from {\em different} events. 
Thus  $\rho_{\mathrm{mix}}(\va )$ does not contain any of the 
dynamical correlation present among particles from the 
same event, but is sensitive to effects determined  
by the single-particle spectra. 
To obtain $\rho_{\mathrm{mix}}(\va )$,
each of the original tracks in an event is replaced  
by a track selected at random from all the other events. 
This is performed after the transformation of  the real events to the
Breit frame.
These ``fake'' events are then used to calculate       
$\rho_{\mathrm{mix}}(\va )$ in exactly the same way as $\rho(\va )$
is calculated from the real events. It should be noted that each fake event 
has, by construction, the same multiplicity ($n$) as the real event, but 
there is no requirement that the fake tracks are drawn from events of 
multiplicity close to $n$. 
This is the same procedure as used by DELPHI \cite{cor_delphi}
and allows direct comparison with theoretical calculations.
However, it has the consequence that the value of $r(\va )$  is influenced  by 
the mixing of events with different multiplicities 
in $\rho_{\mathrm{mix}}(\va )$. 
For this reason many correlation studies, particularly in hadronic
collisions, have used the ``semi-inclusive'' correlation \cite{rew}, 
$r^{(n)}(\va )$, for which both 
$\rho(\va )$ and $\rho_{\mathrm{mix}}(\va )$ are evaluated at 
a fixed multiplicity $n$. 
Monte Carlo studies indicate that, 
for the data studied in this paper, the difference between
$r(\va )$ and $r^{(n)}(\va )$ is quite pronounced 
for low multiplicity ($\la n  (\T ) \ra \leq 2$), 
but largely disappears at high multiplicity. 
            
Analytic calculations for $\hat{r}(\va )$ and $r(\va )$, performed in 
the DLA at asymptotic energies, can be expressed in the form \cite{theo_cor}: 
\beq 
\hat{Y} \equiv -\frac{\ln(\hat{r}(\va )/b)}{2\sqrt{\ln(P\sin\T/\Lambda)}}
\simeq  2 \beta (1-0.5\omega(\va )), 
\label{4}
\eeq
\beq
Y \equiv \frac{\ln r (\va )}{\sqrt{\ln(P\sin\T/\Lambda)}}
\simeq  2 \beta (\omega(\va )-2\sqrt{1-\va }), 
\label{5}
\eeq 
where
\beq
\omega(\va )  
=  2\, \sqrt{1-\va }\, \left(1 -\ln(1-\va)/8\right), 
\label{6}
\eeq
and  $\beta$ and $b$  are given by   
\beq 
\beta = 6 (11\, N_c - 2 n_f)^{-1/2}, 
\qquad
b = 2 \beta \sqrt{\ln(P\sin\Theta/\Lambda)}
\label{7}
\eeq
with $N_c=3$. The effective number of flavours, $n_f$,  
is chosen  to be three  since  the main contribution
to the current-region parton cascade 
comes from light quarks \cite{L-Ochs}. 
These values of $N_c$ and $n_f$ give   
$\beta = 1.15$.

The  analytic QCD calculations 
predict a scaling behaviour, i.e. $\hat{Y}$ and $Y$      
are functions of $\va$ only and depend neither on the initial parton
momentum  $P=Q/2$ nor on  the angle  $\T$.  
According to the calculations, 
both $\YH$ and $\Y$ rise  with  increasing  $\va$.  
The increase of $\YH$ determines 
the behaviour of $\hat{r} (\va )$ which  
is  a strongly decreasing
function due to the decrease in  the number 
of partons (and, hence, parton pairs)
with increasing $\va$ (decreasing $\t$). 
In contrast, $r (\va )$ is predicted to rise  
with increasing $\va$, reflecting 
an increased sensitivity of $r (\va )$ to two-particle
correlations. 
At large  particle separation, i.e. $\t\geq \T$ ($\va \leq 0$),
there are  no two-particle correlations,  
$r (\va \leq 0 ) = 1$ \footnote{This statement follows  
from the analytic DLA calculation, which takes into account 
the angular ordering but neglects momentum conservation 
in the gluon
splitting. On the other hand, Monte Carlo models 
which incorporate momentum conservation exhibit long-range two-particle
correlations in this phase-space region \cite{theo_cor}.}. 

\subsection{Current-Target  correlations}
\label{sub:ct}

A simple way to measure  the  interdependence between 
the number of charged particles in the current and target regions,
$\nc$ and $\nt$,  
is to use the correlation coefficient $\kappa$: 
\beq 
\kappa =\sc^{-1} \st^{-1}\cov ,
\qquad
\cov = \la \nc \nt \ra -  \la \nc \ra \la \nt \ra,
\label{1c}
\eeq
where $\sc$ and $\st$ are  the standard deviations of 
the multiplicity distributions 
in the current and target regions, respectively. 
For positive correlations, $\kappa$  is  positive; for
anti-correlations it is  negative.
The advantages of using definition (\ref{1c}) lie 
in the  simplicity of the boundary conditions,
$-1\leq  \kappa \leq 1$, which allows
a quantitative estimate of the strength of the correlations, and
its low  sensitivity to the losses of particles from the proton remnants, 
which often escape undetected down the beam pipe.
The latter property comes from the fact that  $\cov$ and
the product  $\sc \> \st$ have a similar dependence on the 
total number of measured tracks, 
so that  their ratio has only a small
dependence on the average  track multiplicity in the
target region (see Sect.~\ref{sec:ac}).
 
If hadronisation effects are neglected,
the QCDC and BGF processes lead  to \cite{ch} 
\beq
\cov \simeq  - A_1 (Q^2) R_1 (Q^2, x) - A_2 (Q^2) R_2(Q^2,x), 
\label{cc1}
\eeq
where $A_1(Q^2)$ and $A_2(Q^2)$ are 
$x$-independent functions  
determined by the multiplicity of the outgoing partons.
$R_1 (Q^2, x)$ is the probability for  
back-to-back jet events (i.e. with one jet in the current
and one in the target region)  and $R_2 (Q^2,x)$ is 
the probability   
for an  event  without  jet activity in the current region. 

At small $Q^2$, the values of $A_1 (Q^2)$ and  $A_2 (Q^2)$
are positive \cite{ch}. This leads to a negative covariance (\ref{cc1})
and thus to anti-correlations between the  current and the target regions,  
which can be measured  using eq.~(\ref{1c}).
At asymptotically high $Q^2$, the value of $A_1 (Q^2)$ can be negative. 
In this case the current-target correlations are not  strong and
can even be positive.    

Note that at small fixed $Q^2$, $\cov$  is sensitive to
the BGF rate as a function of $x$ which,  
in turn, depends on the gluon density inside
the proton \cite{ch}. 

\section{Experimental setup}
\label{sec:es}
 
The data were  taken in 1995 at the positron-proton collider HERA
with the ZEUS detector. During this period, the energy of the positron
beam was $E_e=27.5$ GeV  and that of the proton beam was  $820$ GeV.
The integrated luminosity used for the 
present study is 6.4 pb$^{-1}$.
 
ZEUS is a multi-purpose  
detector described in detail in \cite{zeus_det}. 
Of particular importance in the
present study  are the central tracking detector
and the calorimeter.

The central tracking detector (CTD) is  a cylindrical
drift chamber with nine superlayers covering the polar
angle\footnote{ZEUS has a right-handed coordinate system in which
$X=Y=Z=0$ is the nominal interaction point.
The positive $Z$-axis is along the direction of the proton beam.
The $X$-axis is horizontal, pointing towards the centre of HERA.
The polar angle $\theta$ is defined with 
respect to the positive $Z$-direction.} 
region $15^o < \theta < 164^o$ and the
radial range 18.2-79.4 cm. Each superlayer consists
of eight sense wire layers. The transverse momentum resolution
for charged tracks traversing all CTD layers is 
$\sigma(p_{\bot})/p_{\bot} = 0.0058 p_{\bot}  
\oplus 0.0065 \oplus  0.0014/p_{\bot}$ 
with $p_{\bot}$ being the track transverse momentum (in GeV). 
The single hit efficiency of the CTD is greater than $95\%$. 

The CTD
is  surrounded by the uranium-scintillator calorimeter  which is divided 
into three parts: forward (FCAL), barrel (BCAL) and rear (RCAL).
The calorimeter is longitudinally segmented into electromagnetic (EMC)
and hadronic (HAC) sections. The energy resolution of the calorimeter 
under test beam conditions 
is $\sigma_E/E=0.18/\sqrt{E}$ for electrons and 
$\sigma_E/E=0.35/\sqrt{E}$ for hadrons (with $E$ in GeV).

\section{Data selection}
\label{sub:sel}
 
The kinematic variables $x$ and $Q^2$ can  
be reconstructed using  a variety of methods.
In order to determine these variables,  we use 
the measurement of energy and angle of the scattered positron,  
the double angle  and the Jacquet-Blondel  methods \cite{dam}. 
These three methods are further denoted by the subscripts $e$, DA and JB,
respectively. In the double angle method, the 
variables $x$, $Q^2$ and $y$ are reconstructed 
using the angles of the scattered  positron  and the hadronic energy flow.
In the Jacquet-Blondel method, used only in the event selection,
the kinematic variables  are determined entirely from the
hadronic system. The boost vector from the  
laboratory to the Breit frame is determined using the double angle  method,
which is less sensitive to systematic uncertainties in the 
energy measurement than the other methods. 
In the reconstruction of the Breit frame
all charged particles are  assumed to have  pion masses.  

\medskip  
The triggering and online event selections are 
identical to those used in \cite{trigg}. 
To select  neutral-current DIS events  
the following cuts are applied:

\begin{enumerate}
 
\item[$\bullet$]
$E_e^{'}\geq 10$ GeV, $E_e^{'}$ being the energy of the scattered positron.
 
\item[$\bullet$]
$Q^2_{\mathrm{DA}}\geq 10$ GeV$^2$.
 
\item[$\bullet$]
$35 \leq  \delta=\sum E_i(1-\cos\theta_i) \leq 60$ GeV, where
$E_i$ is the energy of the $i^{\mathrm{th}}$  calorimeter cell and 
$\theta_i$ is its polar angle with respect to the beam axis.
 
\item[$\bullet$]
$y_{e} \leq 0.95$.
 
\item[$\bullet$]
$y_{\mathrm{JB}}\geq 0.04$.

\item[$\bullet$]
$Z$-component of the  primary vertex position determined from the   
tracks fitted to the vertex is in the range
$-40 < Z_{\mathrm{vertex}}  < 50$ cm.
 
\item[$\bullet$]
A timing cut requiring that the event time measured by the RCAL
is  consistent with an $\ep$  interaction. 

\item[$\bullet$]
The impact point ($X$, $Y$) of the scattered positron  in the RCAL
has to lie outside a square of $32\times 32$ cm$^2$, 
centred on  the beam axis.
 
\end{enumerate}
 
About 350k   
events  satisfy the above cuts. Tables~\ref{tab:ang} and \ref{tab:fb} 
give the numbers  of selected events for the  kinematic regions
used in this analysis.  
 
The present study is based on the CTD tracks fitted to the vertex.
The scattered positron  was removed from the track sample.
In addition, the following cuts to select tracks 
were imposed:    
 
\begin{enumerate}
 
\item[$\bullet$]
Transverse momentum $p_{\bot}>150$  MeV. 
 
\item[$\bullet$]
$\mid \eta \mid < 1.75$, where $\eta$ is the pseudorapidity
given by $-\ln (\tan(\theta/2))$ with $\theta$ being the polar
angle of the charged track. 
\end{enumerate}

These cuts restrict our
study to  a well-understood, high-acceptance region of the CTD.

To understand
the uncertainties in our results, a subsample of
DIS events containing a single jet was selected   
using calorimeter information. 
The cone algorithm \cite{cone} was used with
radius $R=\sqrt{\delta\eta^2+\delta\phi^2}=1$, where
$\delta\eta$ and $\delta\phi$ are the differences of
pseudorapidity and azimuthal angles of 
energy deposits in the calorimeter
with respect to the jet direction.
Events with  a single jet (in addition to particles in the proton remnant) 
are selected if the jet 
transverse energy $E_T$ is larger than $4$ GeV.
The jet pseudorapidity is restricted to the region
$\mid \eta_{\mathrm{jet}} \mid  < 2$,   
where jets are well reconstructed. 
After the transformation to the Breit frame, 
only jets  which belong  to the current region of the Breit frame
are considered. 

\section{Event simulation}
\label{sec:esim}
 
To determine  the corrections  for selection and acceptance
losses, event migration between $(x, Q^2)$ bins,  and  track
migration between the current and target regions due to
mis-reconstruc\-tion of the Breit frame, a sample of 
neutral-current DIS  events is generated with ARIADNE 4.8 \cite{ARD} 
using tuned  parameters \cite{bruk}.
This  MC is based on the colour-dipole QCD model supplemented 
with the Boson-Gluon
Fusion process. The hadronisation is described by
the JETSET model \cite{je}.
Hadrons with lifetime $c\tau >1$ cm  are treated as  stable.
The GRV94  HO  \cite{GRV}
parameterization of the proton parton distribution functions  is used.
The MC events  obtained  by   this procedure 
define the generator-level sample. 

For  the detector-level sample,
the events are generated with  ARIADNE using the
DJANGO 6.24 \cite{django} program based on HERACLES 4.5.2 \cite{heracles}
in order to  incorporate first-order electroweak corrections.
The events are then  processed  through a simulation of the detector
using GEANT 3.13 \cite{geant} to take into account    
particle interactions with the detector material, particle decays,
event and track migrations, double counting of single tracks,  
resolution, acceptance of the detector and  event
selection.
The detector-level MC events are processed with the same
reconstruction program as used  for the real data.

In addition to ARIADNE,  the LEPTO 6.5 \cite{LEP}
and HERWIG 5.9 \cite{HEW} generators with tuned parameters \cite{bruk} 
are also used to compare with the data. 
The  parton cascade in LEPTO is based on the matrix element calculation
matched to a parton shower (MEPS) according to the DGLAP equation. 
LEPTO orders the parton emissions in invariant mass with an additional
angular constraint to ensure coherence.  
As in ARIADNE, the hadronisation in LEPTO 
is described by the JETSET model.  
HERWIG  also has  a  parton shower based on the DGLAP equation, but 
parton emissions are  ordered in angle. The hadronisation is     
described by a cluster model.   

One of the sources of two-particle correlations  
is the Bose-Einstein interference between identical
particles. By default,  this  
effect is turned off in the JETSET model.  The simulation of
the Bose-Einstein effect is absent in HERWIG.  
According to our studies and 
the DELPHI  results \cite{cor_delphi}, 
the Bose-Einstein interference has a small
effect (less than $2\%$ of relative change) 
on the angular correlations.  

\section{Resolution and acceptance}
\label{sec:ac}

For the present analysis, it is important to understand  the two-particle
resolution of the CTD. The resolution is determined in the  
Breit frame as the variance of the absolute difference 
between $\t$ measured at the generator-level  and 
the detector-level Monte Carlo  defined above.
It is  found that to ensure that the two particles are resolved,  the
angle $\t$ between two tracks  in the current region of the Breit frame
should not be smaller than two  degrees.
For a given $\T$, $P$ and $\Lambda$, this limit  determines  the maximum
values  of $\va$ used in  this analysis.

For the cuts used, the  event and current-region
track acceptances   are  both  in the range  $70 - 90\%$, depending on
the $(Q^2, x)$  region studied. This is sufficiently good to
measure the angular correlations reliably.

For the target region,
the track acceptance lies in the range $20-30\%$. 
The low acceptance results  mainly from the
$\mid \eta \mid < 1.75$ restriction.
Using the ARIADNE model, the low target-region 
acceptance is  found to have a small effect on the
current-target correlations measured according to (\ref{1c}): 
the restriction  $\mid \eta \mid < 1.75$ decreases
the absolute value of $\cov$  by  $50\%$. However, 
this decrease is compensated partially in $\kappa$ by a similar  
decrease in the standard deviations of eq.~(\ref{1c}).
As a result, 
the cut  $\mid \eta \mid < 1.75$ decreases
the absolute value of $\kappa$ by only $15\%$.

If the average charged multiplicity $\la \nt \ra$
is measured in the target region as a 
function of the multiplicity $\nc$ in the
current region, the restriction $\mid \eta \mid < 1.75$ decreases
the $\la \nt \ra$ by a factor of $2$ to $5$, 
depending on $\nc$. Therefore,
for the present  target-region acceptance, 
the average number of particles
in the target region as a function of the
multiplicity in the current  region (or  vice versa)
cannot be reliably determined, which is why $\kappa$ is used,  
as described in  subsection~\ref{sub:ct}.

\section{Uncorrected observables}
\label{sec:unc}

To understand how well the MC models describe 
simple observables related to the angular and current-target
correlations, 
we compare the uncorrected data to the MC distributions 
after the detector simulation.
Table~\ref{tab:ang} gives the numbers  of
events in  each region  of $Q^2$, the average values of $Q^2$ and
the average charged multiplicity  in
the current region of the Breit frame for the data.

\begin{figure}
\begin{center}
\mbox{\psfig{file=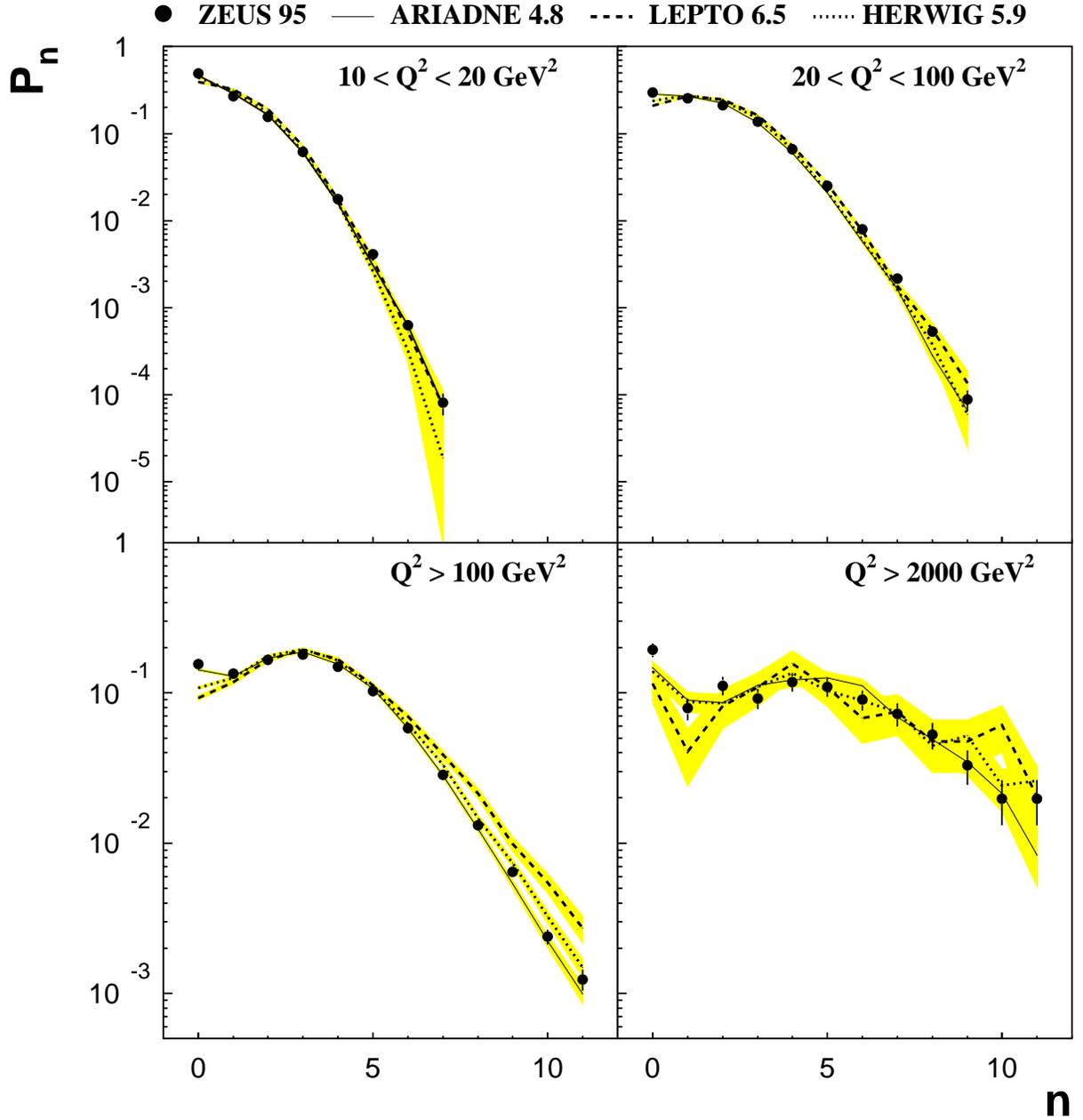, height=18.0cm}}
\caption{
Probability  distribution $P_n$ for  detecting $n$ 
charged particles in the current region of the Breit frame
for different ranges of $Q^2$. The uncorrected data are  compared to  
MC predictions after the detector simulation.
The errors on the data are  the statistical
uncertainties. The shaded bands show the statistical 
uncertainties on the MC predictions. Table~\ref{tab:chi} 
gives the $\chi^2/$NDF of the MC predictions.} 
\label{fig:01}
\end{center}
\end{figure}

Fig.~\ref{fig:01} shows the probability, $P_n$, 
of detecting $n$ charged particles in the current 
region of the Breit frame for four  $Q^2$ regions. 
The difference between the data and Monte Carlo models  
is illustrated by the  $\chi^2/\mathrm{NDF}$
in Table~\ref{tab:chi}.
Note that, although  the  angular 
correlations are sensitive to $P_n$,
they are  also determined by the semi-inclusive   
two-particle densities \cite{rew}.  
  
\begin{figure}
\begin{center}
\mbox{\epsfig{file=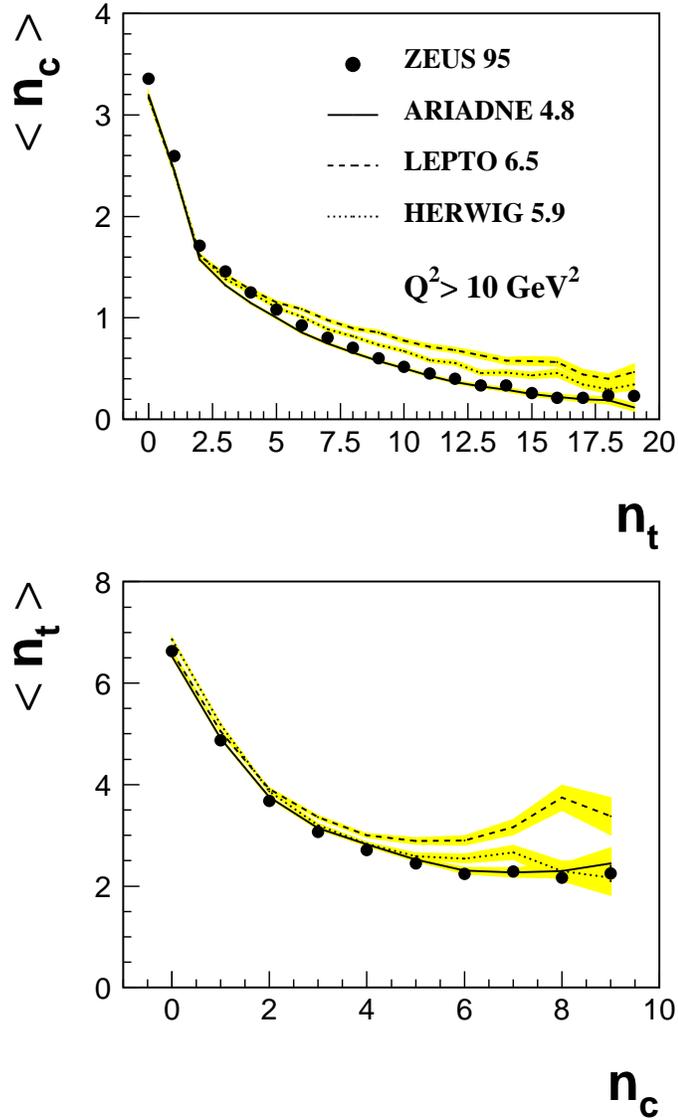, height=18.0cm}}
\caption{
Average charged multiplicities in the
current (target) hemisphere as a function of the
multiplicity in the opposite hemisphere for $Q^2 > 10$ GeV$^2$.
The uncorrected data are  compared to
MC predictions after the detector simulation.
The statistical uncertainties  on the data are  
typically smaller than the size of the symbols. 
The shaded bands show the statistical uncertainties  on the MC predictions.} 
\label{fig:02}
\end{center}
\end{figure}

The average charged multiplicities in the
current (target) hemisphere as a function of the
multiplicity in the opposite hemisphere are shown 
for $Q^2>10$ GeV$^2$ in Fig.~\ref{fig:02}.
In both  cases the average multiplicity in a hemisphere 
decreases as the number of particles  in 
the opposite hemisphere increases,
which is  a  signature of anti-correlations 
between current and target charged multiplicities.

These comparisons show that ARIADNE gives the best description of the
uncorrected observables.
Therefore, this model is used to correct the correlations  
for detector effects.    

\section{Correction procedure }
\label{sec:cp}

Due to the complexity of the measured variables, 
a bin-by-bin correction procedure is used. 
The correction factors $\mathcal{C}$ for  each kinematic region
in ($Q^2$, $x$) are   evaluated separately for each
observable, $\mathcal{A}=\hat{r}(\va ), r(\va ), \kappa $,
\beq 
\mathcal{C} =
\frac{\mathcal{A}^{\mathrm{gen}}}{\mathcal{A}^{\mathrm{det}}}, 
\label{cor1} 
\eeq
where $\mathcal{A}^{\mathrm{gen}}$ is  calculated 
at  the generator-level of ARIADNE and  $\mathcal{A}^{\mathrm{det}}$ is
that at  the detector-level of this model.
The corrected value for an observable is  found by multiplying
its measured value by the relevant  correction factor.  

The correction
factors  are  close to unity for all observables  and
vary smoothly for any given $\mathcal{A}$.   
For $\hat{r} (\va )$ and $r (\va )$,
the correction factors are  never larger  than $1.3$  
and decrease with increasing $\va$.  
For the  current-target correlations, the correction factors  never
exceed $1.35$. The sign of these  correlations
remains  unchanged after the correction procedure.
 
\section{Statistical and systematic uncertainties}
\label{sec:ss}

The main sources of systematic uncertainties are:  

\begin{enumerate} 

\item[$\bullet$]
Event reconstruction and selection. The systematic
check was  performed by varying the cuts on 
$y_{e}$, $y_{\mathrm{JB}}$, $\delta$
and the vertex position requirement 
($y_{e} \leq  0.90$, $y_{\mathrm{JB}} \geq  0.05$,
$40 \leq  \delta \leq  55$ GeV,  $-35 < Z_{\mathrm{vertex}}  < 45$ cm).
The contribution of these uncertainties for the angular and current-target
correlations is  typically $60\%$ of the total systematic  
errors.

\item[$\bullet$] 
Track reconstruction and selection.
The cuts were tightened:
tracks should have transverse momenta larger than $200$  MeV
and $\mid \eta \mid < 1.5$.  
Tracks which reach at least the 
third CTD  superlayer were  used.
These  uncertainties  are typically $30\%$ 
of the total systematic errors. 

\item[$\bullet$]
The use of tracks not
fitted to the primary vertex.

\end{enumerate}

The overall systematic uncertainty is  determined by adding 
the  uncertainties discussed above in quadrature.
The error bars on  
the corrected data presented in Sect.~\ref{sec:results} include both
statistical and systematic errors, added in quadrature.
The errors on the angular correlations
are dominated  by the  statistical  errors
(about $80 -90\%$ of the total error).
For the current-target correlations,
the major uncertainty comes from the systematic effects.

No systematic uncertainty is attributed to the use of LEPTO or HERWIG  
in place of ARIADNE to determine the correction factors since these
models do not adequately describe the raw data
for several variables 
relevant to the correlation study.
If the difference between the correction
factors determined using ARIADNE and those using LEPTO or HERWIG 
is added, the systematic errors  increase by about $50\%$.

\section{Results}
\label{sec:results}

\subsection{Two-Particle angular correlations}

\begin{figure}
\begin{center}
\mbox{\epsfig{file=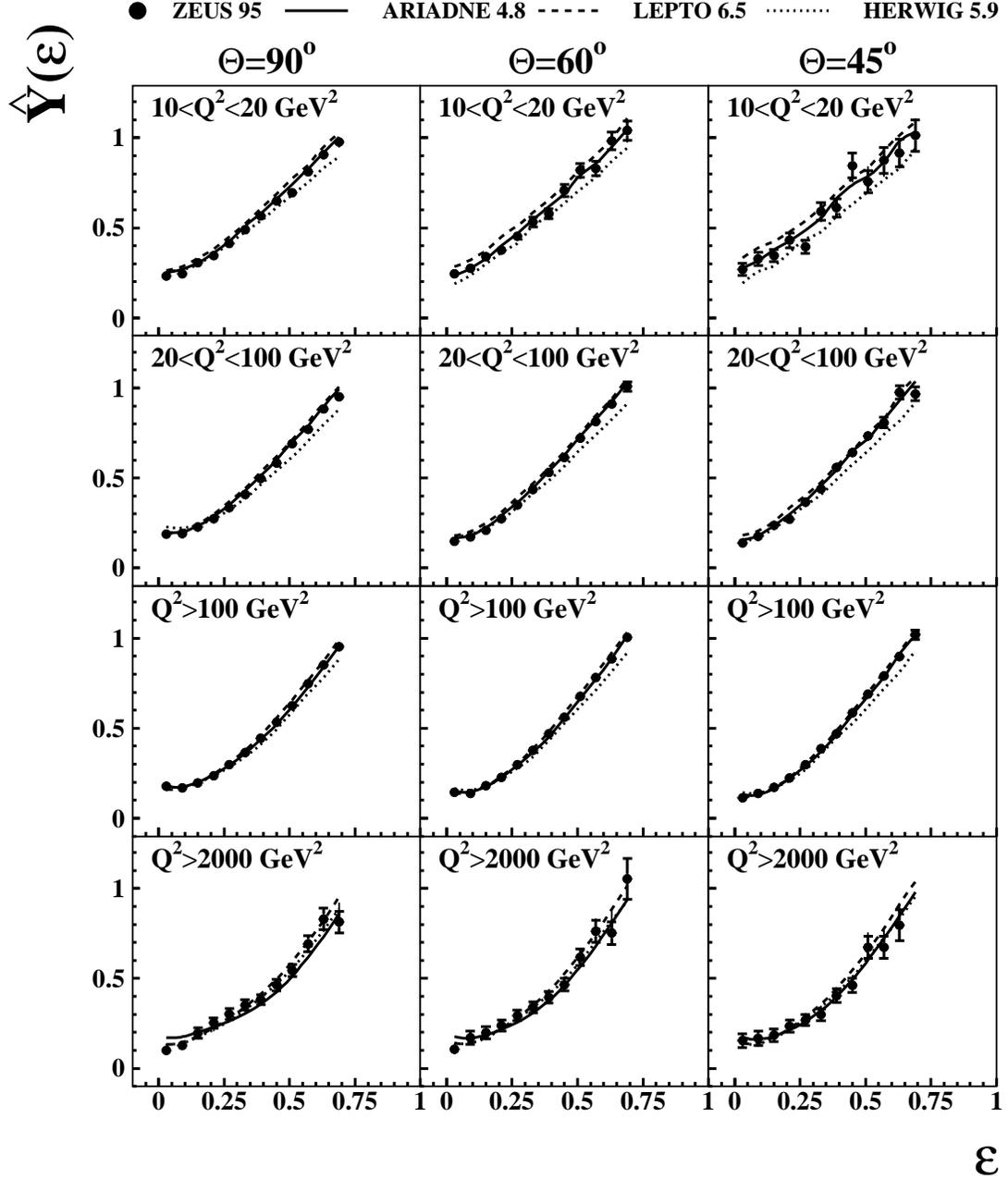, height=18.0cm}}
\caption{ 
The normalised particle density $\YH$ in the current region
for four $Q^2$ ranges and three different $\Theta$ values 
compared to
MC predictions. The rescaling is performed using
$\Lambda=0.15$ GeV.     
The inner bars on the corrected data show the statistical uncertainties.
The full error bars include the systematic
uncertainties, which are typically negligible compared to the
statistical errors.}
\label{fig:2}
\end{center}
\end{figure}

\begin{figure}
\begin{center}
\mbox{\epsfig{file=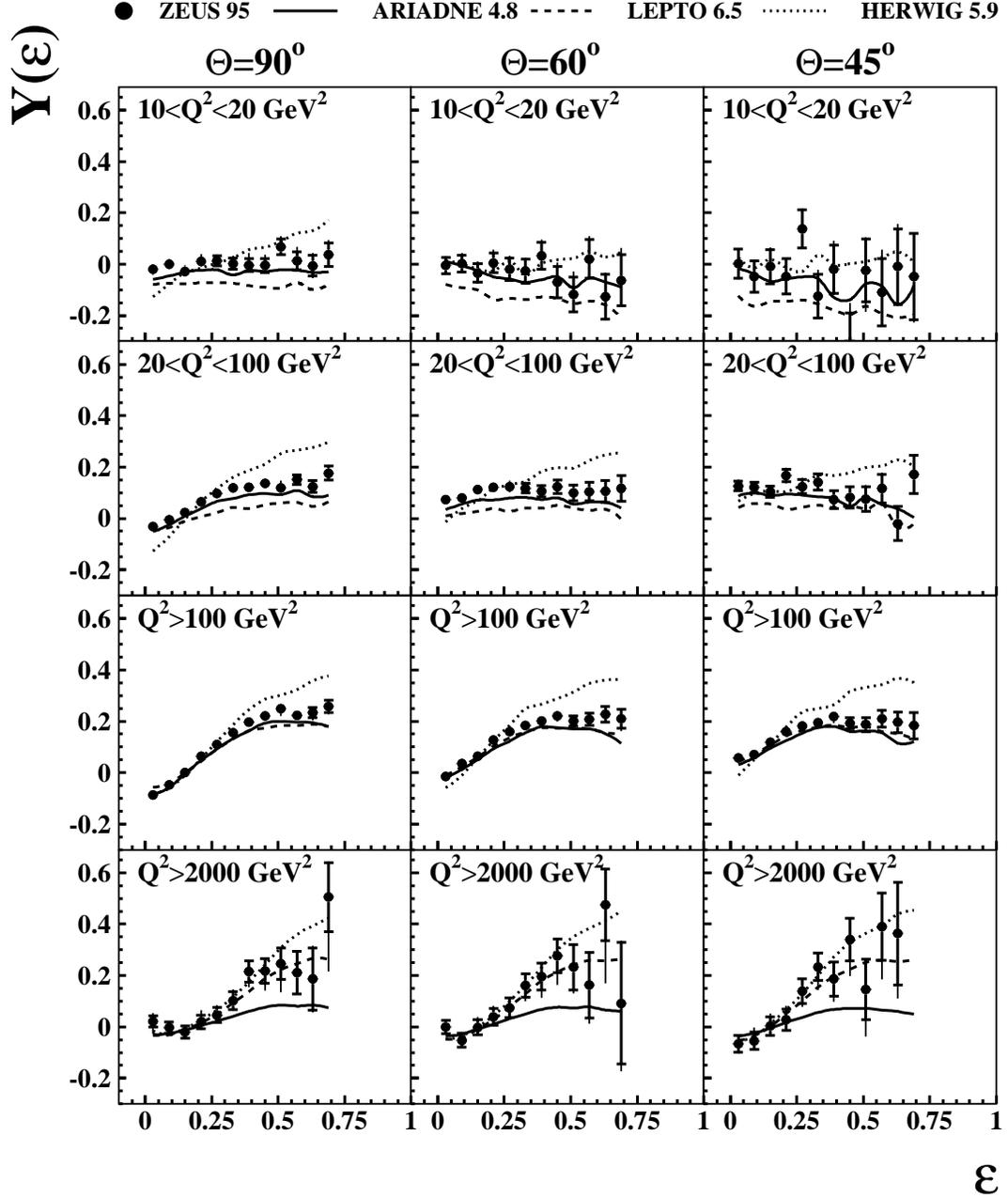, height=18.0cm}}
\caption{ 
The correlation function $\Y$  in the current region 
for four $Q^2$ ranges and three different $\T$ values
compared to MC predictions. 
The rescaling is performed using
$\Lambda=0.15$ GeV. 
The inner error bars on the corrected data show the statistical uncertainties
and the full error bars  include the systematic uncertainties.} 
\label{fig:3}
\end{center}
\end{figure}

The behaviour of the normalised particle density, 
$\YH$, (see eq.~(\ref{4}))  
and correlation function, $\Y$, (see eq.~(\ref{5})), 
measured in the four  $Q^2$ regions, listed in Table~\ref{tab:ang},
are shown in Figs.~\ref{fig:2} and ~\ref{fig:3} for three  values of $\T$.
For these  figures, the value of $\Lambda=0.15$ GeV is 
chosen in  order to be consistent with 
similar measurements made at LEP \cite{cor_delphi,fluc_l3}.

The values of $\YH$ increase with increasing  $\va$,
reflecting  a strong decrease of the number of
particle pairs with increase of  $\va$. 
Such a trend is mostly determined by the single-particle distribution,
rather than correlations between particles.  
In contrast, $\Y$ is more sensitive  to  
two-particle correlations. It  rises  with increasing $\va$ 
at large $Q^2$  where the particle  multiplicity is larger  
and the jet structure is more pronounced.
This behaviour implies an increase in the strength of  
the correlations with decreasing angular separation between
the two particles. At low  $Q^2$, however, this  rise is not observed 
due to the small charged particle multiplicity in
the current region.

The $\YH$ distributions of the MC models  
agree   reasonably well with the data.   
The ARIADNE model describes the data in all $Q^2$ regions.
At low $Q^2$ LEPTO slightly overestimates $\YH$, while
HERWIG underestimates it.  
The agreement is less satisfactory   for $\Y$. 
ARIADNE  is quite  successful in the description
of low and medium $Q^2$, but underestimates the data
at high $Q^2$. LEPTO fails to describe the data 
at $Q^2\le 100$ GeV$^2$. HERWIG gives 
a poor description of $\Y$ in all $Q^2$ regions, except 
for the highest $Q^2$.
 
In the current version of ARIADNE 4.8, the suppression of available phase
space for parton radiation due to 
the extended nature of the target remnant can
also affect the current region of the Breit frame at high $Q^2$. 
A recent (as yet unreleased) modification of  ARIADNE \cite{plom},
which ensures that  the whole of the available current-region phase space is
used for gluon radiation, gives a good description of $\Y$ at high
$Q^2$ (not shown).  

\begin{figure}
\begin{center}
\mbox{\epsfig{file=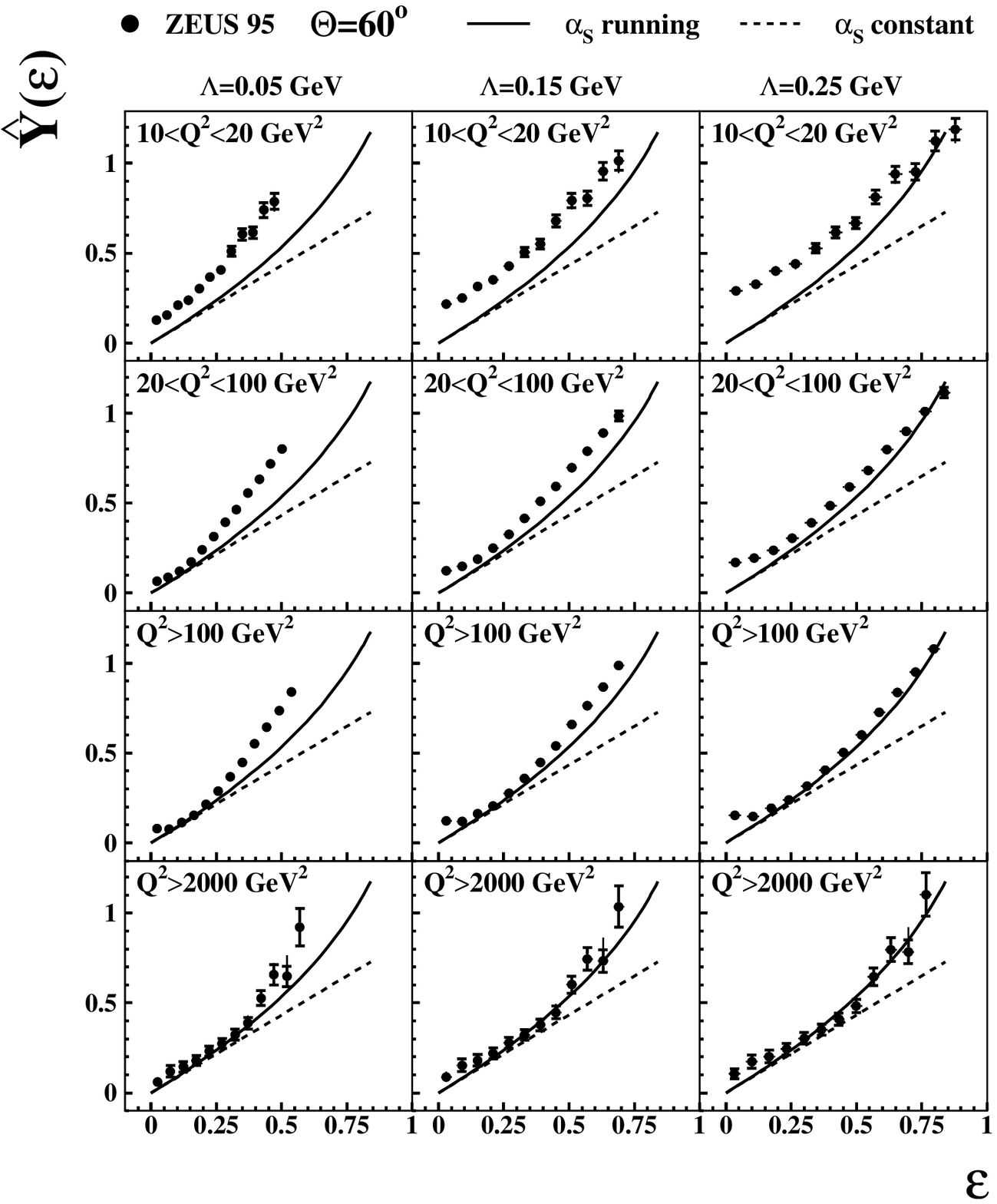, height=18.0cm}}
\caption{ 
The normalised particle density 
$\YH$ in the current region for $\T = 60^o$ compared to
the analytic QCD predictions (\ref{4}) for different 
effective $\Lambda$ values.
Solid (dashed) lines show the predictions at asymptotic energy for
a running (fixed) coupling constant.} 
\label{fig:4}
\end{center}
\end{figure}

The results indicate that there are no  
changes in the $\YH$ and $\Y$ distributions  
at large $\va$ as the angle $\Theta$ is varied  
in the range from $45^o$ to $90^o$. This observation supports the 
predicted scaling property of   
eqs.~(\ref{4}) and (\ref{5}).  To compare  
the analytic calculations with the data,
we choose $\Theta = 60^o$. Fig.~\ref{fig:4} shows  $\YH$ 
and the asymptotic  QCD predictions (\ref{4}) 
for fixed (dashed lines) and running (solid lines)
strong coupling constant\footnote{
An evolution of the parton shower  at a {\em fixed} 
energy scale $Q^2$ is characterized
by a running coupling constant $\alpha_s$ which 
reflects  a change of energy scale as the parton shower evolves.} 
for three different values of $\Lambda$. 
For the calculations with a fixed coupling constant, 
the values $\Lambda=0.05, 0.15, 0.25$ GeV 
for $Q^2>2000$ GeV$^2$ correspond to
the effective $\alpha_s=0.107, 0.130, 0.144$ for
the lowest order QCD relation between the $\alpha_s$ and $\Lambda$. 
The assumption of a fixed coupling constant is inconsistent with the
data for all $Q^2$ intervals, while the calculation with
a running coupling constant shows
reasonable  agreement  with the 
data at $Q^2>2000$ GeV$^2$  and large $\Lambda$. 

\begin{figure}
\begin{center}
\mbox{\epsfig{file=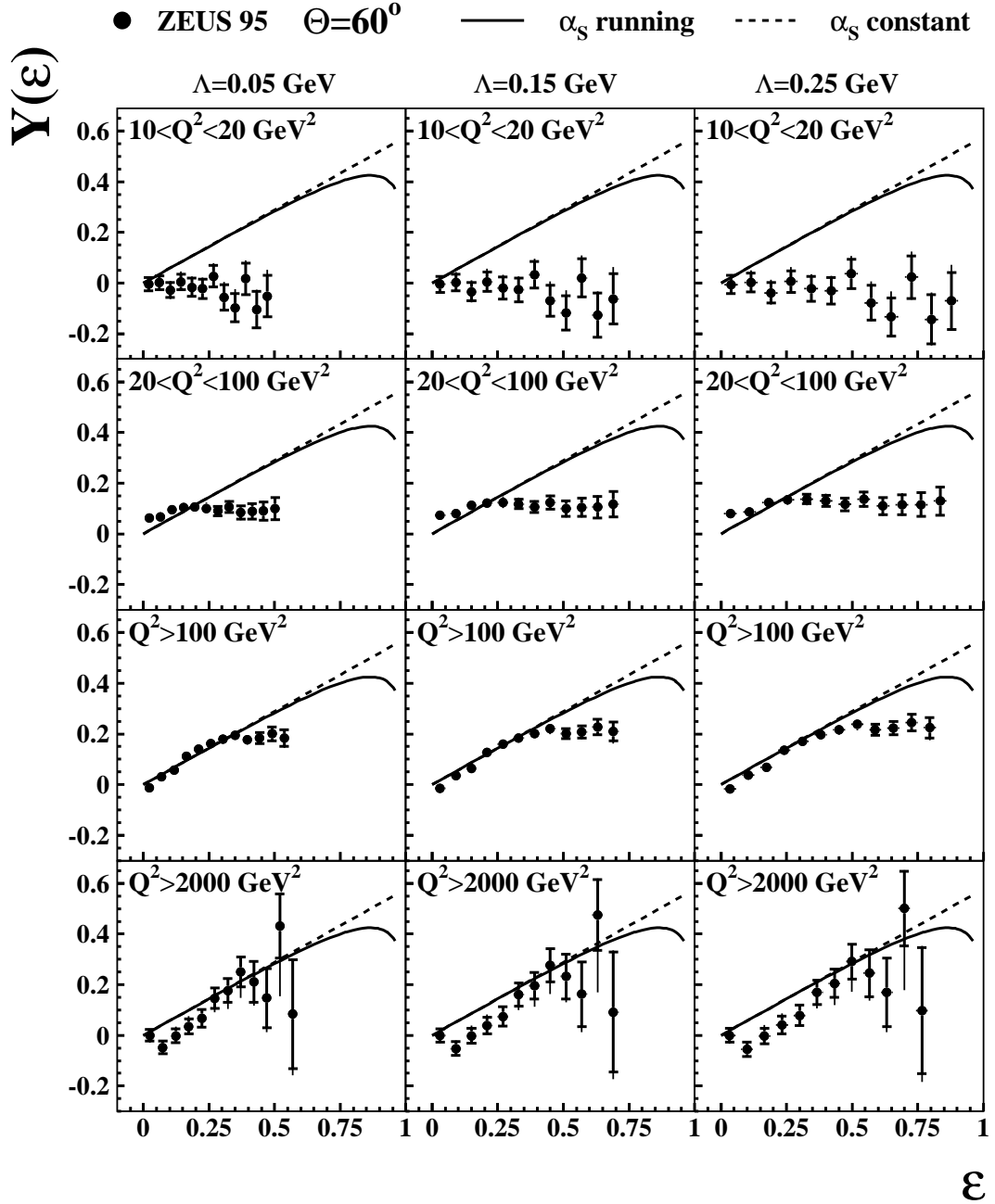, height=18.0cm}}
\caption{ 
The correlation function $\Y$ in the current region 
for $\T = 60^o$  compared to
the analytic QCD predictions (\ref{5}) for different 
effective $\Lambda$ values.
Solid (dashed) lines show the predictions at asymptotic  energy for
a running (fixed) coupling constant.}
\label{fig:5}
\end{center}
\end{figure}

A similar comparison of 
the analytic calculations for $\Y$ with the data
is shown in Fig.~\ref{fig:5}. As for  $\YH$, discrepancies are 
observed   at  small $Q^2$. The agreement is better at high $Q^2$; 
however, the calculations  still overestimate  the data at small $\va$.  
This discrepancy is likely to be related to the neglect of
energy-momentum conservation in the DLA, as recently 
discussed in \cite{cor_delphi,fluc_l3}. 
This simplification  makes the angular correlations in
the analytic calculations  
more prominent for large angular separations, $\theta_{12}$ (small $\va$). 

\begin{figure}
\begin{center}
\mbox{\epsfig{file=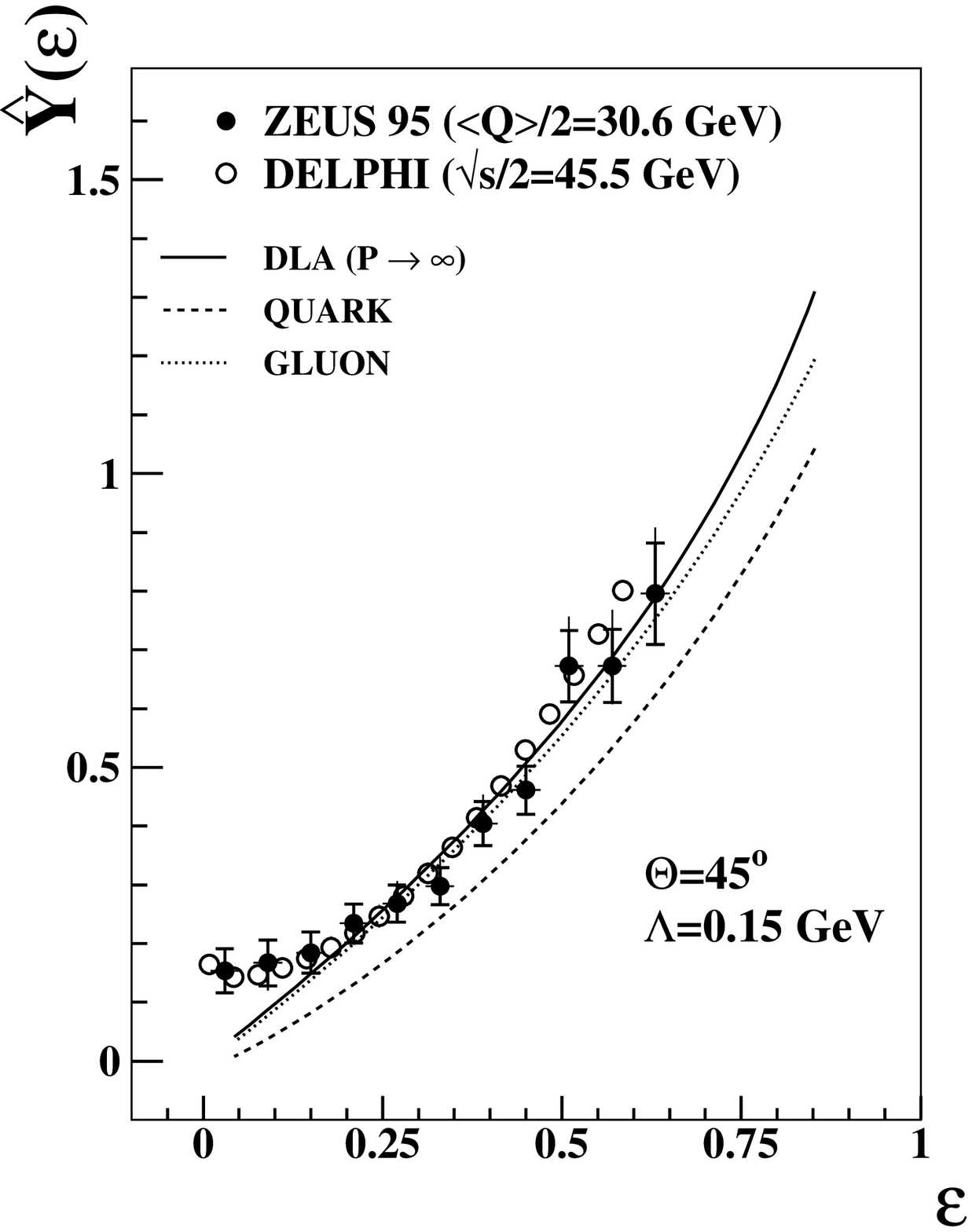, height=17.0cm}}
\caption{ 
The normalised particle density $\YH$
for $Q^2>2000$ GeV$^2$ ($P=\la Q\ra/2 =30.6$ GeV) compared to
the analytic QCD predictions and 
DELPHI results for  $P=45.5$ GeV. For
both experiments  $\T = 45^o$ and  $\Lambda=0.15$ GeV. 
The solid line 
shows the DLA predictions at asymptotic  energy 
for a running coupling constant. 
The dashed  (dotted) line corresponds to the  prediction at finite energy 
($P=30.6$ GeV) for quark (gluon) jets.}  
\label{fig:5a}
\end{center}
\end{figure}

\begin{figure}
\begin{center}
\mbox{\epsfig{file=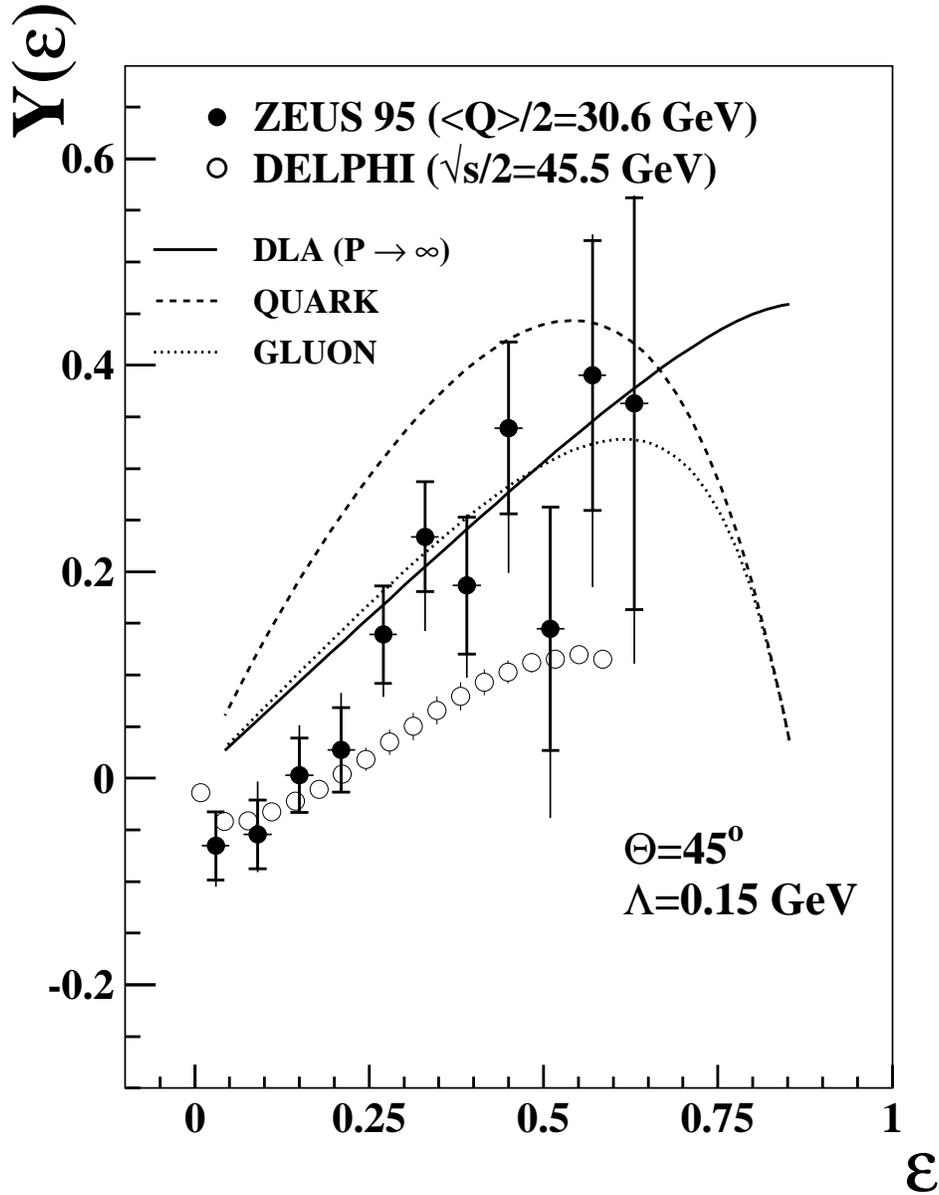, height=17.0cm}}
\caption{ 
The correlation function $\Y$ 
for $Q^2>2000$ GeV$^2$ ($P=\la Q\ra/2 =30.6$ GeV)  compared to
the analytic QCD predictions and
DELPHI results for  $P=45.5$ GeV.
For both experiments $\T = 45^o$ and  $\Lambda=0.15$ GeV. 
The solid line shows the DLA predictions at asymptotic  
energy for a running coupling constant.
The dashed  (dotted) line corresponds to the  prediction at finite energy 
($P=30.6$ GeV) for quark (gluon) jets.} 
\label{fig:5b}
\end{center}
\end{figure}

Figs.~\ref{fig:5a} and ~\ref{fig:5b} show our results for 
$\YH$ and $\Y$ at $\Theta=45^o$ and $\Lambda =0.15$ GeV 
together with the DELPHI data at $P=45.5$ GeV 
($\sqrt{s}=91$ GeV) \cite{cor_delphi}. 
The ZEUS data are shown for $Q^2 > 2000$ GeV$^2$, which  
corresponds approximately to $P=\la Q\ra/2=30.6$ GeV.  
Despite the difference in energy, the results for $\YH$
agree.  Note that no significant energy dependence of 
$\YH$ was found at LEP \cite{cor_delphi}.

According to \cite{cor_delphi}, the behaviour of
$\Y$ at $\sqrt{s}=183$ GeV is steeper
than  that at $\sqrt{s}=91$ GeV. If there is a universality 
between the current region of DIS and a single hemisphere of
$\E$, one might  expect that
$\ep$ data at $P=\la Q\ra/2=30.6$ GeV should  exhibit
a less strong rise than $\E$ data at $P=45.5$ GeV.  
However, the ZEUS data shown in Fig.~\ref{fig:5b} exhibit 
the opposite trend, which suggests differences between   
the angular correlations in  the current region of DIS 
and a single hemisphere of $\E$ annihilation. 
The observed discrepancy  
might  be related to different choices of the axis of the $\Theta$-cone:
whereas this axis in $\E$ was  determined by the sphericity axis,
the virtual photon direction is used to 
determine the axis of the $\Theta$-cone in DIS. 
In addition, the  leading-order  QCD effects
discussed in Sect.~\ref{sec:meth} 
can lead to an uncertainty in our results.
To investigate this issue, similar studies of the angular 
correlations were performed using only single-jet events. 
The single-jet pre-selection  described in Sect.~\ref{sub:sel} 
leads to smaller values  of $\Y$ than the data 
shown  in  Fig.~\ref{fig:5b}, 
but a small discrepancy with $\E$ still remains.

Several further checks have been performed to understand  
the difference  between the ZEUS and the DELPHI results 
for $\Y$. Firstly,  
$\ln(P\T /\Lambda)$ was used in the denominator of (\ref{1}), 
as in the DELPHI analysis.
Secondly, a small fraction of events 
at $Q^2>8000$ GeV$^2$ was rejected, so  that
no event has an initial parton at 
an energy higher than $P=45.5$ GeV.   
The analysis was repeated by calculating $\va$ with
$P=\sqrt{\la Q^2 \ra} /2$, rather than using $P=Q/2$ for
each individual event. In addition,   
HERWIG and the modified version of ARIADNE  
without the suppression effect in the current region were used.  
Both models describe $\Y$ at $Q^2 > 2000$ GeV$^2$,
and, therefore, they are better suited  to correct the data 
in this region. For the checks discussed above, 
no significant changes in the ZEUS data were  observed
which can account for the discrepancy.

Figs.~\ref{fig:5a} and \ref{fig:5b} 
also show the analytic QCD results \cite{theo_cor}
at infinite energy and finite energy ($P=30.6$ GeV) 
separately for quark and gluon jets. 
The analytic prediction for gluon jets 
shows  better agreement with the data than that for quark jets. 
However, it  again fails  to describe $\Y$  at  small $\va$. 
A  possible source of the discrepancy with the
finite-energy quark prediction is 
the ratio $R=9/4$ of the mean parton
multiplicity in gluon and quark jets used 
in the calculations for quark jets.  
If  one goes beyond
the DLA, the value of this ratio is smaller than 9/4.
For example, in the next-to-leading log approximation, 
$R\simeq 1.8$ \cite{RAT}.
A smaller  value of $R$
will bring the analytic prediction for quark
jets  closer  to the gluon prediction \cite{WOchs}. 

\subsection{Current-Target correlations}

\begin{figure}
\begin{center}
\mbox{\epsfig{file=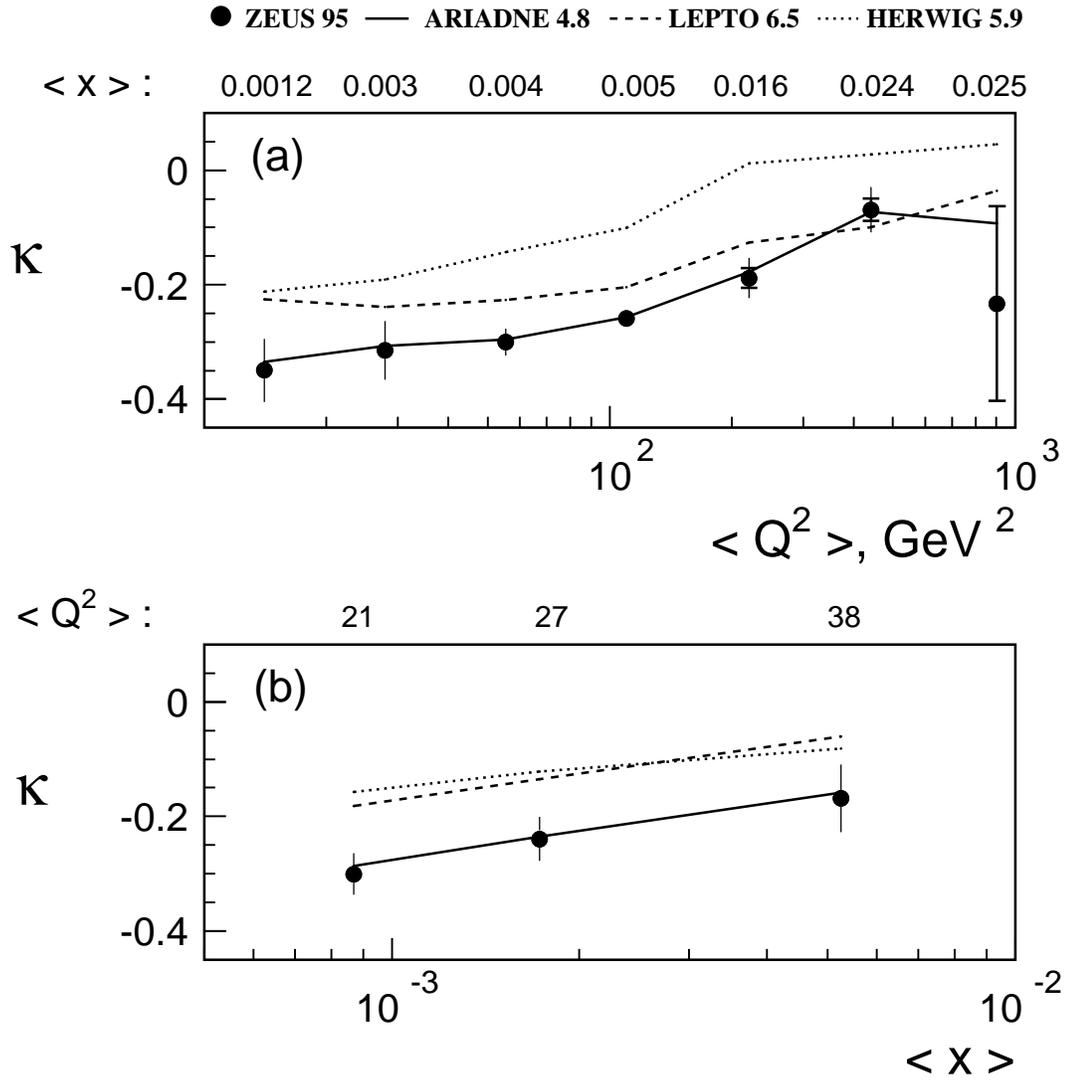, height=18.0cm}}
\caption{ 
{\bf (a)} represents the  evolution of  the 
coefficient of correlations $\kappa$
with predominant variation in  $Q^2$
for corrected data and MC predictions; 
{\bf (b)} shows the same quantity where predominantly $x$ varies. 
The  corrected values of  
$\la Q^2\ra$ and $\la x \ra$ are indicated for each plot.    
The inner error bars on the data show the statistical uncertainties.  
The full error bars  include the systematic uncertainties. } 
\label{fig:rho}
\end{center}
\end{figure}

Figure~\ref{fig:rho}
shows the behaviour of  the correlation coefficient
$\kappa$ as a function of the average values of $Q^2$ and  $x$. 
The same bins in $Q^2$ and $x$ were used as  
in  previous studies  \cite{zeusFF, zeusM}. 
Instead of calculating $\kappa$  in  each bin,
the bins are combined to increase the statistics.
Table~\ref{tab:fb} gives the numbers of events 
and  the  uncorrected average  values of $Q^2$ and $x$. 
Fig.~\ref{fig:rho}a 
shows the dependence of the correlations  on  $Q^2$, 
while  Fig.~\ref{fig:rho}b shows  the  $x$ variation.   
Note that the average values of  $Q^2$ and $x$ shown  
in the figure are corrected.

Anti-correlations ($\kappa < 0$) are  observed for all values 
of $x$ and $Q^2$, as predicted by eq.~(\ref{cc1}). 
The  magnitude  of $\kappa$ 
decreases  with increasing  $\la Q^2 \ra $
from $0.35$ to $0.1$.  
According to the analytic result of eq.~(\ref{cc1}), 
the observed anti-correlations  can be due to   
the $\cal{O}$$(\as )$ effects (QCDC and BGF). Their kinematics
in the Breit frame can reduce the particle multiplicity  in the
current region and increase it in the target region.   
A Monte Carlo study \cite{ch} shows that the contribution  from 
hadronisation is relatively small.    

The ZEUS result, shown in Fig.~\ref{fig:rho},  
gives a quantitative estimate of the
correlations  between the current and target region multiplicities
in DIS at HERA, an  effect  which  is  
quite different from that observed between 
the forward and backward hemisphers in  
$\E$ annihilation \cite{ee,del,al,op}. 
The  kinematics of the $\cal{O}$$(\as )$ effects 
in the Breit frame which lead to these
correlations are also a possible source of 
the disagreement  between ZEUS and $\E$ data 
for the angular correlations. These effects can also lead 
to discrepancies  between the data and the analytic QCD 
calculations which do not take into account the kinematics
of the $\cal{O}$$(\as )$ processes in the Breit frame. 
It is noteworthy that  the current region 
of DIS  shows a different  average multiplicity
at small $Q^2$  than a single  hemisphere of $\E$ annihilation
\cite{h1FF,zeusM,zeusMN} due to the $\cal{O}$$(\as )$ processes.

The magnitude of the anti-correlations  increases
with decreasing $\la x \ra$. 
According to the analytic result (\ref{cc1}), 
this can be due to an increase of the fraction of events  with one or 
two jets produced in the target region. This behaviour is driven by 
an increase of the gluon density inside the
proton, leading to an increase of the Boson-Gluon Fusion rate.

The ARIADNE model agrees  well with the data.
A reasonable  description of dijet 
production  in DIS by this model \cite{dijet} 
is possibly responsible  for this agreement. 
The LEPTO and HERWIG predictions show  the same  trend but
do not reproduce the magnitude of the correlations which  
is likely to be related to a lower  predicted dijet rate  \cite{dijet}.
 
\section{Conclusions}

The evolution of  two-particle angular correlations 
with $Q^2$ of DIS has been  studied in the Breit frame. 
The data have been compared to the results of numerical
simulations and to those of analytic QCD calculations in the DLA for
partons, assuming  the Local Parton-Hadron Duality hypothesis.

The results on the angular correlations support  the predicted scaling
behaviour in the angle $\Theta$ at large $\va$. 
The scaling in energy is found to hold approximately for $\YH$.
For this variable, our results are 
in good agreement with those of DELPHI for $\E$,
despite the difference in the energy. 
In contrast to $\YH$, the  two-particle
correlation function $\Y$ depends  strongly on $Q^2$:  
at  low $Q^2$, the two-particle correlations are suppressed, but   
they increase with $\va$ at high $Q^2$ ($Q^2>100$ GeV$^2$). 

The analytic calculations for $\YH$ and  $\Y$ do not 
describe the data at low $Q^2$. 
The results become closer to the data as $Q^2$ increases. 
The asymptotic QCD predictions with a running
coupling constant  only  describe the $\YH$  
at $Q^2 > 100$ GeV$^2$ for
large values of the  effective $\Lambda$.
For the highest $Q^2$ studied, the predictions
reproduce the trend of $\Y$  
but overestimate the data at small $\va$. 

The best description of the angular correlations
is achieved by the ARIADNE model, although it fails to
describe $\Y$ at the highest  $Q^2$ studied.  
The prediction of ARIADNE shows
smaller correlations in this region than the data  
due to the suppression of available current-region phase
space by the remnant in the colour-dipole model.
LEPTO and HERWIG show small
discrepancies for $\YH$.  For the $\Y$ distribution, 
which is more sensitive
to the two-particle correlations, both  models fail 
to describe the data at low $Q^2$. 

The current-target correlations in the Breit frame 
are found to be large and  negative. 
They are thus very different from those measured
in $\E$ collisions, where  
small and positive forward-backward correlations have been
observed.  In DIS, the  correlations    
increase with  decreasing  $x$. 
Such a behaviour is  expected from an  
increase of the Boson-Gluon Fusion
rate that leads to an increase of dijet production.
ARIADNE  agrees  well with the current-target correlations.
Neither LEPTO nor  HERWIG  reproduce the 
magnitude of the correlations. Insufficient dijet production
in these models is  a possible source of  this failure. 

Many studies at HERA have found similarity between   
multiparticle production in  
the current region of $\ep$ collisions and a single
hemisphere of $\E$ annihilation. 
This paper demonstrates that correlations are a powerful
tool for investigating this area.
For high $Q^2$, the ZEUS results show  
differences between the two-particle angular correlations 
in DIS and $\E$ annihilation; this is inconsistent
with the universality of two-particle inclusive densities,  
which are expected to be sensitive to the kinematics
of the first-order QCD processes producing final-state  hadrons
in  both the current and target regions. 

\section*{Acknowledgments}
This paper was completed shortly after the tragic and untimely
death of Prof. Dr.~B.~H.~Wiik. All members of the ZEUS Collaboration
wish to acknowledge the tremendous role which he played in
the success of the HERA project and this experiment.
His leadership and friendship will be greatly missed.

We wish to express our gratitude to the DESY  accelerator division for
the excellent performance of the HERA  machine.  We acknowledge the
effort of all engineers and technicians who have participated in the
construction and maintenance of the ZEUS  experiment.
The support and encouragement of the DESY directorate continues to be
invaluable for the work of the ZEUS collaboration.
We thank B.~Buschbeck, W.~Kittel, W.~Metzger, W.~Ochs and  J.~Wosiek   
for helpful  discussions.

\newpage
\clearpage
\pagebreak
 
{}

\newpage 

\begin{table}
\begin{center}
\begin{tabular}{|c|c|c|c|}
\hline
$Q^2$ (GeV$^{2}$) range & $\la Q^2\ra$ GeV$^{2}$ & 
$\la n \ra$ &No. of events \\ 
\hline\hline 
   10-20  & $14.4\pm 0.5$ & $0.96\pm 0.04$  & 161421 \\
\hline
20-100  & $40\pm 1$  & $1.5\pm  0.1$ & 159453 \\
\hline
  $>$ 100  & $320\pm 11$ & $2.9\pm 0.3$ & 31461 \\
\hline
  $>$ 2000  & $3737\pm 203$ & $3.9\pm 0.6$ & 456 \\
\hline
\end{tabular}
\end{center}
\caption{Numbers of selected events used  
to study the  angular correlations in  different $Q^2$ regions,
integrated over all $x$ values. The corresponding uncorrected average 
values of $Q^2$ and average charged multiplicity $\la n\ra$ 
in the current region of the Breit frame are also shown. The errors 
are the combined statistical and systematic uncertainties.} 
\label{tab:ang}
\end{table}

\begin{table}
\begin{center}
\begin{tabular}{|c|c|c|c|c|}
\hline
$Q^2$ (GeV$^2$) range & $x$ range & 
$\la Q^2 \ra$ GeV$^{2}$ & $\la x\ra$ & No. of events
\\
\hline
\hline
\multicolumn{5}{|c|}{evolution with $Q^2$}  \\
\hline
\hline
   10-20  & $0.6-2.4\cdot 10^{-3}$ & $14.2\pm 0.6$ & $(1.22\pm 0.01)\cdot 10^{-3}$  & 79381  \\
\hline
   20-40  & $1.2-10\cdot 10^{-3}$  & $28.7\pm 0.7$ & $(2.88\pm 0.01)\cdot 10^{-3}$ & 48287   \\
\hline
  40-80   & $1.2-10\cdot 10^{-3}$  & $55\pm 2$ & $(3.82\pm 0.01)\cdot 10^{-3}$ & 34956  \\
\hline
 80-160   & $2.4-10\cdot 10^{-3}$  & $109\pm 3$ &  $(5.29\pm 0.02)\cdot 10^{-3}$  &  11331  \\
\hline
160-320   & $2.4-50\cdot 10^{-3}$  & $218\pm 5$ &  $(2.1\pm 0.1)\cdot 10^{-2}$  & 4350  \\ 
\hline
320-640   & $10-50\cdot 10^{-3}$  & $416\pm 9$ &  $(2.4\pm0.2)\cdot 10^{-2}$ &  1615  \\ 
\hline
640-1280   & $10-50\cdot 10^{-3}$  & $899\pm 14$ &  $(1.7\pm 0.2)\cdot 10^{-2}$  &   270  \\ 
\hline
\hline
\multicolumn{5}{|c|}{evolution with $x$}  \\ 
\hline
\hline
10-1280 & $0.6-1.2\cdot 10^{-3}$ & $20.7\pm 0.4$ & $(8.7\pm 0.1)\cdot 10^{-4}$ 
& 72039 \\
\hline
10-1280 & $1.2-2.4\cdot 10^{-3}$ & $28\pm 2$ & $(1.7\pm 0.1)\cdot 10^{-3}$ 
& 74621 \\
\hline
10-1280 & $2.4-10\cdot 10^{-3}$ & $53\pm 5$ & $(4.4\pm 0.1)\cdot 10^{-3}$  
& 73825 \\
\hline
\end{tabular}
\end{center}
\caption{Bins in $Q^2$ and $x$ used to study 
the current-target correlations.
The average values of $Q^2$ and $x$ are 
shown without detector corrections. The errors 
are the combined statistical and systematic uncertainties.}
\label{tab:fb}
\end{table}

\begin{table}
\begin{center}
\begin{tabular}[b]{|c|*{3}{r@{.}l|}}
\hline
\multicolumn{1}{|c|}{$Q^2$ (GeV$^{2}$)  range }
& \multicolumn{6}{c|}{$\chi^2$/NDF} \\
\cline{2-7}
& \multicolumn{2}{c|}{ARIADNE 4.8} & \multicolumn{2}{c|}{LEPTO 6.5}
& \multicolumn{2}{c|}{HERWIG 5.9}  \\
\hline
\hline
 10-20 & \hspace{0.8cm} 25&{4} & \hspace{0.6cm} 189&{6} & \hspace{0.6cm} 73&{8} \\
\hline
20-100  &  \hspace{0.8cm} 27&{7}  & \hspace{0.6cm} 167&{9}  & \hspace{0.6cm} 71&{1} \\
\hline
$>$ 100  &  \hspace{0.6cm} 4&{5}  & \hspace{0.4cm} 34&{3}  & \hspace{0.6cm} 32&{9} \\
\hline
$>$ 2000  &  \hspace{0.6cm} 1&{0} & \hspace{0.6cm} 1&{3}  & \hspace{0.0cm} 0&{9} \\
\hline
\end{tabular}
\end{center}
\caption{
$\chi^2 =
\sum_i \left(P_n^{\mathrm{Data}}-P_n^{\mathrm{MC}}\right)^2/
(\sigma_{\mathrm{Data}}^2+
\sigma_{\mathrm{MC}}^2)$ per degree of freedom
for the uncorrected probability distribution $P_n$ shown
in Fig.~\ref{fig:01}. }
\label{tab:chi}
\end{table}


\newpage 

\newpage

\newpage

\newpage 

\newpage 

\newpage

\newpage

\newpage

\newpage

\newpage

\end{document}